\documentclass[twocolumn,showpacs,preprintnumbers,amsmath,amssymb,superscriptaddress,pral,aps,longbibliography]{revtex4-2}

\usepackage{graphicx}
\usepackage{xcolor}
\usepackage[hidelinks]{hyperref}
\hypersetup{
    colorlinks,
    linkcolor={black},
    citecolor={blue},
    urlcolor={blue!80!black}
}

\allowdisplaybreaks

\begin{document}

\title{Few-to-many-particle crossover of pair excitations in a superfluid}

\author{Fabian Resare}
\affiliation{Department of Physics, Chalmers University of Technology, 41296 Gothenburg, Sweden}

\author{Johannes Hofmann}
\email{johannes.hofmann@physics.gu.se}
\affiliation{Department of Physics, Gothenburg University, 41296 Gothenburg, Sweden}
\affiliation{Nordita, Stockholm University and KTH Royal Institute of Technology, 10691 Stockholm, Sweden}

\date{\today}

\begin{abstract}
Motivated by recent advances in the creation of few-body atomic Fermi gases with attractive interactions, we study theoretically the few-to-many-particle crossover of pair excitations, which for large particle numbers evolve into a mode that describes amplitude fluctuations of the superfluid order parameter (the ``Higgs'' mode). Our analysis is based on the hypothesis that salient aspects of the excitation spectrum are captured by  interactions between time-reversed pair states in a harmonic oscillator potential. Microscopically, this assumption leads to a Richardson-type pairing model, which is integrable and thus allows a systematic quantitative study of the few-to-many-particle crossover with only minor numerical effort. We first establish a parity effect in the ground-state energy, i.e., a spectral convexity in the energy of open-shell configurations compared to their closed-shell neighbors, which is quantified by a so-called Matveev-Larkin parameter discussed for mesoscopic superconductors, which generalizes the pairing gap to mesoscopic ensembles and which behaves quantitatively differently in a few-body and a many-body regime. The crossover point for this quantity is widely tunable as a function of interaction strength. We then compute the excitation spectrum and demonstrate that the pair excitation energy shows a minimum that deepens with increasing particle number and shifts to smaller interaction strengths, consistent with the finite-size precursor of a quantum phase transition to a superfluid state. We extract a critical finite-size scaling exponent that characterizes the decrease of the gap with increasing particle number.
\end{abstract}

\maketitle

A fundamental question in physics is how the constituents of interacting systems behave collectively to yield the emergent properties of matter~\cite{anderson72}: How many particles are required to transition from a few-body ensemble to a collective many-body state? Experiments that explore this crossover are scarce: Examples include the emergence of metallic behavior in colloidal clusters~\cite{rademann87}, semiconductor quantum dots~\cite{alivisatos96,reimann02}, or superconductivity in nanograins~\cite{ralph95,black96,ralph97}. Recent experiments on few-body ensembles of ultracold fermions provide a new platform in which the few-to-many-body crossover can be studied~\cite{serwane11,zuern13}. These cold atom setups offer several advantages over solid-state systems: For example, they are free of disorder or impurities, and parameters such as the trap geometry and the interaction strength are controlled very accurately. Moreover, precise measurement techniques exist, such as modulation spectroscopy~\cite{bayha20} or even direct fluorescence imagining of the few-body wave function~\cite{holten21a,holten22}. On the theoretical side, it is a significant challenge to describe the few-to-many-particle crossover, with most theoretical methods tailored to describe either few-particle ensembles or the many-particle limit.

In this Letter, we show how the excitation spectrum of a paired superfluid emerges at large particle numbers, with a Higgs mode that describes amplitude fluctuations of the superfluid order parameter. The key theoretical advance that allows us to interpolate between the few- and many-particle limit is the use of an integrable pairing model, which treats the interaction between time-reversed states exactly. Corresponding experiments in the few-body limit were recently carried out in two-dimensional harmonically trapped atomic Fermi gases~\cite{bayha20}, which create pair excitations on top of a few-body ground state by modulating the interaction strength. For particular closed-shell configurations, these excitations transfer an atom pair to a higher oscillator shell and are hence gapped. With increasing attractive interaction, however, the excitation energy decreases below the noninteracting value and assumes a minimum before it increases again at strong interactions. The expectation is that in the limit of large particle numbers, the minimum shifts to weaker interactions and decreases to zero energy, signaling a quantum phase transition to a superfluid state~\cite{bruun14}. Hence, the observations~\cite{bayha20} are interpreted as the few-body precursor of a quantum phase transition. Here, we corroborate this picture with exact results beyond the few-body limit, allowing in particular the extraction of a finite-size exponent for the gap closing at large particle number.

In order to highlight the crucial role of pair interactions, we consider a model [stated in Eq.~\eqref{eq:hamiltonian} below] with an effective interaction between time-reversed fermion pairs and show that it quantitatively describes current experiments. 
Similar pairing models are able to account for pairing fluctuation effects on ground state properties of interacting one-dimensional Fermi systems~\cite{hofmann16,hofmann17}, and we show here that they are also capture the excitation spectrum of two-dimensional interacting fermions. 
The model belongs to a class of so-called Richardson models~\cite{richardson63,richardson64,richardson65,richardson66}, which are solved by an algebraic Bethe ansatz (see Refs.~\cite{vondelft01,dukelsky04,johnson23} for reviews). This allows for an efficient calculation of many-body states at only minor numerical effort compared to other methods such as a full-configuration-interaction approach~\cite{rontani09,sowinski13,damico15,sowinski15,grining15,bjerlin16,rontani17,bekassy22,bekassy24}, stochastic variational methods~\cite{laird24}, coupled-cluster methods~\cite{grining15}, or Monte Carlo calculations~\cite{berger15,rammelmueller16,luo16}. The advantage of the pairing model is thus twofold: First, it links the observed excitation spectrum to a microscopic BCS-type pairing principle in mesoscopic systems. Second, being integrable, calculations are very tractable and can be performed both for few-particle and many-particle ensembles.

We study the following Hamiltonian:
\begin{equation}
H = \sum_{j} \varepsilon_j n_j - g \sum_{ij} A_i^\dagger A_j^{} . \label{eq:hamiltonian}
\end{equation}
Here, $\varepsilon_j$ is a single-particle energy labeled by a primary quantum number $j$, and the energy level is $d_j$-fold degenerate with degenerate states distinguished by a second quantum number $m$. For a two-dimensional (2D) harmonic oscillator, for example, we have $\varepsilon_j = j+1$ ($j=0,1,\ldots$)  (all energies are expressed in units of the harmonic oscillator energy) with $d_j = j+1$ per spin and $m=-j,-j+2,\ldots,j$ is the angular momentum projection within a shell $j$. Furthermore, we define the particle number operator in a given shell
\begin{equation}
n_{j} = \sum_m (c_{jm}^\dagger c_{jm}^{} + c_{j\bar{m}}^\dagger c_{j\bar{m}}^{}) ,
\end{equation}
where $c_{jm}^\dagger$ creates a fermion in a state $(j,m)$ and $c_{j\bar{m}}^\dagger$ creates a fermion in a corresponding time-reversed state (i.e., with opposite spin and angular momentum). The interaction term couples pairs of fermions annihilated by
\begin{equation}
A_j = \sum_m c_{j\bar{m}}^{} c_{jm}^{} .
\end{equation}
We briefly comment on the link with a contact interaction: In occupation number representation, the contact interaction contains a matrix element~\cite{hofmann16}
\begin{align}
&w_{(j_1,m_1),(j_2.m_2),(j_3,m_3),(j_4,m_4)} \nonumber \\[1ex]
&= \int d{\bf r} \, \phi_{j_1,m_1}^*({\bf r}) \phi_{j_2,m_2}^*({\bf r}) \phi_{j_3,m_3}({\bf r}) \phi_{j_4,m_4}({\bf r}) . \label{eq:wijkl}
\end{align}
where $\phi_{j,m}({\bf r})$ is the single-particle wave function of the state with quantum number $(j,m)$. 
For an oscillator potential, where Eq.~\eqref{eq:wijkl} conserves the total angular momentum, the effective integrable interaction term in Eq.~\eqref{eq:hamiltonian} neglects some matrix elements with pair-breaking transitions compared to a contact interaction~\cite{hofmann16} (and will thus not reproduce certain features of the contact interaction such as a nonrelativistic conformal symmetry at weak interactions~\cite{bekassy22,bekassy24}), but as we will demonstrate it accurately describes the pair excitation spectrum. Note that for a box potential, where single-particle wave functions are plane-wave states labeled by a two-dimensional wave vector ${\bf k}$, the pairing interaction~\eqref{eq:hamiltonian} connects pairs with opposite momenta and spin, \mbox{$({\bf k}, \uparrow; - {\bf k},\downarrow) \to ({\bf k}', {\uparrow;} -{\bf k}',\downarrow)$}, which are precisely the matrix elements that give rise to superfluidity~\cite{matveev97}. In the present case, the interaction~\eqref{eq:hamiltonian} connects pairs in time-reversed states of the 2D harmonic oscillator, \mbox{$(j, m; j,\bar{m}) \to (j', m'; j',\overline{m'})$}.

Fermionic systems with pairing interactions as in Eq.~\eqref{eq:hamiltonian} are integrable~\cite{richardson63,richardson64,richardson65,richardson66}. A key observation for the solution is that the interaction only couples empty and pair-occupied levels while singly occupied levels decouple from the Fock space. Restricted to a set ${\cal U}$ of pair levels, Eq.~\eqref{eq:hamiltonian} then describes a quadratic Bose Hamiltonian with a hard-core constraint on $A_j$ characterized by $[A_j, A_i^\dagger] = \delta_{ij} (d_j - n_j)$. This constraint can be incorporated into the Bose gas solution~\cite{vondelft00}: Interacting eigenstates evolve continuously from noninteracting states and are thus described by a set of $M$ occupied pair levels \mbox{$\{(j_0,m_0),\ldots,(j_{M-1},m_{M-1})\} \subset {\cal U}$} (the dependence on the time-reversed index is implied here). The corresponding state
\begin{equation}
|(j_0,m_0),\ldots,(j_{M-1},m_{M-1})\rangle = B_{0}^\dagger \ldots B_{{M-1}}^\dagger |0\rangle  \label{eq:state2}
\end{equation}
is created by acting on the empty state $|0\rangle$ with the same creation operators as for the Bose problem
\begin{equation}
B_\nu^\dagger = \sum_{\ell=0}^{L} \frac{A_\ell^\dagger}{2 \varepsilon_{\ell} - E_\nu} . \label{eq:BJ}
\end{equation}
Here, the $E_\nu$ are generalized pair level energies (called roots), which are the central objects in the Richardson solution, and $L$ is a cutoff on the single-particle spectrum. In particular, the energy of the state~\eqref{eq:state2} is 
\begin{equation}
E = \sum_{\nu=0}^{M-1} E_\nu \label{eq:energy}
\end{equation}
with additional single-particle contributions of blocked levels. The state~\eqref{eq:state2} is an eigenstate of Eq.~\eqref{eq:hamiltonian} provided that the roots $E_\nu$ solve the Richardson equations
\begin{equation}
\frac{1}{g} - \sum_{\ell\in {\cal U}}\frac{d_\ell}{2 \varepsilon_{\ell} - E_{\nu}} + \sum_{\mu=0, \mu\neq\nu}^{M-1} \frac{2}{E_{\mu} - E_{\nu}} = 0 . \label{eq:richardson}
\end{equation}
Here, the first two terms are the same as for noninteracting bosons and the last term, which couples different occupied levels, comes from the hard-core constraint on~$A_j$. 

%++++++++++++++++++++++++++++++++++++++++
\begin{figure}[t!]
\includegraphics{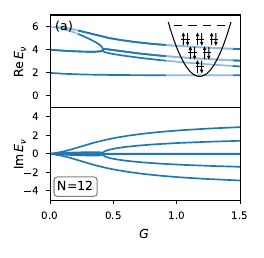}\hspace{-0.4cm}
\includegraphics{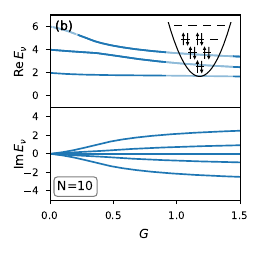}
\vspace{-0.3cm}
\caption{Evolution of the Richardson roots with increasing attractive pair interaction for (a) a closed-shell configuration with \mbox{$N=12$} particles and (b) an open-shell configuration with \mbox{$N=10$} particles in a 2D harmonic trap.}
\label{fig:1}
\end{figure}
%++++++++++++++++++++++++++++++++++++++++

We solve the Richardson equations~\eqref{eq:richardson} iteratively starting from a noninteracting configuration with a set of blocked and pair-occupied levels, the latter setting the initial conditions for the roots $E_\nu$ (for alternative solution methods, see Refs.~\cite{faribault11,elaraby12,johnson23}).  Throughout the Letter, to make contact with the experiment~\cite{bayha20}, we consider a 2D harmonic oscillator with a cutoff at twice the Fermi level of an $N$-particle ground state configuration. We subtract a linear divergence in the second term of the Richardson equations by defining a renormalized coupling 
\begin{equation}
\frac{1}{g} = \frac{1}{G} + \sum_{\ell=0}^L \frac{d_\ell}{2 \varepsilon_{\ell}} .
\end{equation}
Roots derived from degenerate levels are split in the complex plane, and with increasing interaction strength there can be singular points at which two or more roots merge or bifurcate. These singularities are not reflected in the energy~\eqref{eq:energy} or other observables~\cite{claeys15}, but they need to be resolved by the solution algorithm (see Ref.~\cite{resare22} for a detailed discussion of the numerical implementation). 

We first examine the ground state configurations: To illustrate the Richardson solution, Fig.~\ref{fig:1}(a) shows the roots $E_\nu$ as a function of the interaction strength for $N=12$ particles, which has a singular point near $G = 0.5$ (here involving the merging of three roots). The inset depicts the corresponding noninteracting parent state from which the state evolves. (Note that the pairing Hamiltonian couples states within a shell equally, hence the ordering of pair- or singly-occupied states within a shell is not important.) The parent configuration here is nondegenerate  with completely filled orbitals (for the 2D harmonic oscillator, this happens for so-called ``magic'' numbers \mbox{$N=2,6,12,20,\ldots$}). For comparison, Fig.~\ref{fig:1}(b) shows an \mbox{$N=10$} configuration with an open shell at the Fermi level, where degenerate roots are split in the complex plane without further merging.

%++++++++++++++++++++++++++++++++++++++++
\begin{figure}[t!]
\includegraphics{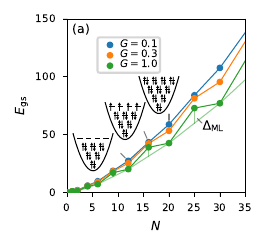}\hspace{-0.4cm}
\includegraphics{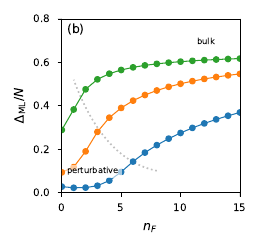}\vspace{-0.3cm}
\caption{(a) Ground-state energy for several interaction strengths $G=0.1,0.3$, and $1$ (blue, orange, and green points). The thin green lines added to the \mbox{$G=1$} results offer a geometric interpretation of the Matveev-Larkin parameter $\Delta_{\rm ML}$, Eq.~\eqref{eq:matveevlarkin}, as the excess energy relative to the mean of neighboring closed-shell configurations. (b) Matveev-Larkin parameter~\eqref{eq:matveevlarkin} for the same interaction strengths as in (a) as a function of Fermi level $n_F$. The dashed gray line indicates the transition region between the perturbative and the bulk result and is drawn to guide the eye.}
\label{fig:2}
\end{figure}
%++++++++++++++++++++++++++++++++++++++++

Figure~\ref{fig:2}(a) shows the ground-state energy as a function of particle number for three different interaction strengths $G=0.1,0.3$, and $1$. Here, we consider configurations with nondegenerate ground states, corresponding to the closed-shell configurations discussed above as well as open-shell configurations in which all states at the Fermi level are filled with one spin species (see the inset for an illustration). As is apparent from the figure, closed-shell states are more strongly paired compared to their open-shell neighbors, which is a general feature of attractive pairing interactions~\cite{lee07} that we here generalize to degenerate systems. The excess energy of open-shell configurations is~\cite{matveev97}
\begin{equation}
\Delta_{\rm ML} = E_{\rm gs}^{\rm open}({n_F}) - \frac{1}{2} \Bigl[E_{\rm gs}^{\rm cl}(n_F-1) + E_{\rm gs}^{\rm cl}(n_F)\Bigr] . 
\label{eq:matveevlarkin}
\end{equation}
Here, \mbox{$E_{\rm gs}^{\rm open}(n_F)$} and \mbox{$E_{\rm gs}^{\rm cl}(n_F)$} are open-shell and closed-shell ground-state energies with Fermi level $n_F$, respectively. In this notation, for example, \mbox{$E_{\rm gs}^{\rm cl}(n_F=3)$} and \mbox{$E_{\rm gs}^{\rm cl}(n_F=4)$} are the ground-state energies of closed-shell configurations with \mbox{$N=12$} and \mbox{$N=20$} particles, respectively [left and right insets in Fig.~\ref{fig:2}(a)], and \mbox{$E_{\rm gs}^{\rm open}({n_F}=4)$} is the energy of an open-shell configuration with \mbox{$N=16$} particles where the valence energy levels are simply occupied [middle inset in Fig.~\ref{fig:2}(a)]. 
The thin green lines, which connect the ground-state energies for full shells, offer a geometric interpretation of the excess energy defined in Eq.~\eqref{eq:matveevlarkin}: The second term in that equation is the average of the neighboring closed-shell energies, which lies on the thin green line evaluated at the open-shell particle number. The excess energy is then the difference between this value and the open-shell energy, and it is indicated by the vertical green line in Fig.~\ref{fig:2}(a). 
The quantity~\eqref{eq:matveevlarkin} is called the Matveev-Larkin parameter~\cite{matveev97} and it generalizes the bulk gap $\Delta$ to mesoscopic ensembles: Indeed, for large particle number, it is (up to finite-size corrections)~\cite{matveev97,roman03}
\begin{equation}
\Delta_{\rm ML} = d_{n_F} \Delta ,
\end{equation}
where $\Delta$ solves the gap equation,
\begin{equation}
\frac{1}{G} = \sum_{\ell=0}^L \frac{d_\ell}{2 E_\ell} \label{eq:gapequation}
\end{equation}
with \mbox{$E_\ell = \sqrt{(\varepsilon_\ell - \mu)^2+\Delta^2}$} and $\mu$ is a chemical potential that constrains \mbox{$N=\sum_\ell [1-(\varepsilon_\ell-\mu)/E_\ell]$}. For the harmonic oscillator, \mbox{$\Delta \sim \sqrt{N}$} for large $N$ such that the Matveev-Larkin parameter is extensive. By contrast, for small particle numbers and interaction strength, its value is set by the perturbative result
\begin{equation}
\Delta_{\rm ML} = d_{n_F} \frac{G}{2} ,
\end{equation}
with $d_{n_F}$ the degeneracy of the Fermi level, which scales as ${\it O}(\sqrt{N})$ with a much smaller magnitude.
 Indeed, a full solution of the pairing model agrees with this picture as shown in Fig.~\ref{fig:2}(b), which shows the intensive quantity $\Delta_{\rm ML}/N$ for the same interaction strengths as in Fig.~\ref{fig:2}(a). The crossover between the perturbative and the bulk regime is most apparent for \mbox{$G=0.1$}, and the crossover region shifts to a smaller particle number with increasing interaction (indicated by the dotted line). Note that even a small increase in the interaction strength has a strong effect on the crossover region compared to 1D systems~\cite{hofmann17}, which we attribute to the increased density of states in 2D traps. While existing experiments on one-dimensional quantum wires~\cite{cheng15} or trapped atoms~\cite{zuern13} appear firmly in the perturbative limit~\cite{hofmann16,hofmann17}, our calculations indicate that experiments on 2D Fermi gases may allow to explore this crossover problem in full even for moderate particle numbers.

%++++++++++++++++++++++++++++++++++++++++
\begin{figure}[t!]
\includegraphics{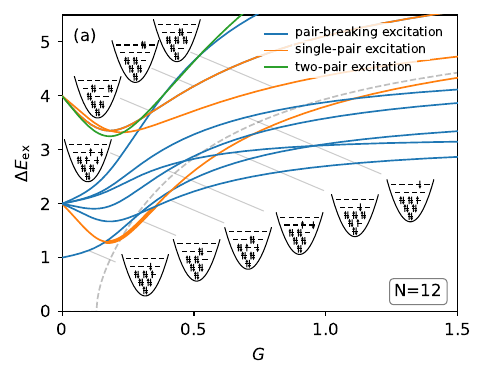}\hspace{0.5cm}
\includegraphics{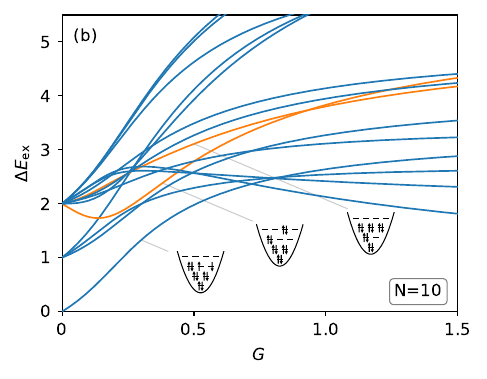}
\caption{Excitation spectrum relative to the ground-state energy $\Delta E_{\rm ex}$ for (a) \mbox{$N=12$} and (b) \mbox{$N=10$} particles in a harmonic trap as a function of interaction strength, where pair-breaking excitations are shown up to the second level. We include the parent state configuration for several excitations. The gray dashed line in (a) is the mean-field prediction $2\Delta$ for the Higgs excitation.}
\label{fig:3}
\end{figure}
%++++++++++++++++++++++++++++++++++++++++

We proceed to discuss excited states, which are the main results of the Letter. Figure~\ref{fig:3}(a) shows the excitation spectrum, where $\Delta E_{\rm ex}$ denotes the  energy of excited states relative to the ground-state energy, for \mbox{$N=12$} particles, which has a closed-shell ground state, where we include as an inset with each excitation branch the occupation of the noninteracting parent state. Excitations are obtained either by creating pair excitations or by breaking pairs and promoting single fermions to higher levels, where we show pair-breaking excitations only up to the second level for clarity. 
At weak interactions, excitation energies have a linear perturbative interaction shift, which is predominantly positive for pair-breaking excitations and negative for pair excitations. 
With increasing interaction strength, pair excitations show an avoided level crossing with the ground state, where the excitation energy develops a minimum and increases again at larger coupling. This is clearly visible in the figure, which shows one pair excitation starting at energy $2$ (orange line) and additional single-pair and two-pair excitations starting at $4$ (orange and green lines). Importantly, at intermediate interactions, the lowest pair excitation is the first excited state (bold orange highlight), a result that holds for all closed-shell configurations studied. For comparison, we show in Fig.~\ref{fig:3}(b) the excitation spectrum of an open-shell $N=10$ state. Here, the lowest excitation evolves from the degenerate noninteracting state with two single fermions in blocked states at the Fermi level. 
Note that the structure (and counting) of excited states with pair and single-particle excitations shown in the figures extends straightforwardly to larger ensembles. 

%++++++++++++++++++++++++++++++++++++++++
\begin{figure}[t!]
\includegraphics{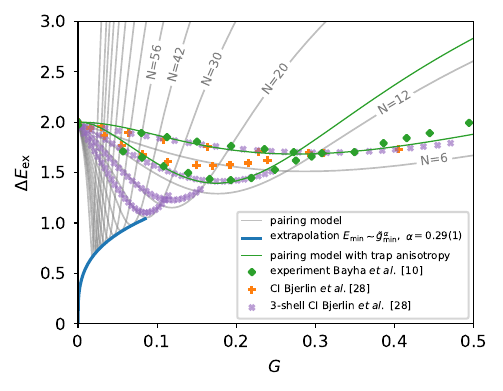}
\caption{Lowest pair excitation energy relative to the ground state $\Delta E_{\rm ex}$ for the first $15$ closed-shell configurations. The blue line extrapolates the excitation minimum to large particle number.  Green lines take into account a small harmonic trap anisotropy to match experimental results (green diamonds) of Ref.~\cite{bayha20}. Orange and purple crosses show full-configuration approach calculations by Ref.~\cite{bjerlin16}.}
\label{fig:4}
\end{figure}
%++++++++++++++++++++++++++++++++++++++++

The minimum of the first pair excitation is interpreted as the few-body precursor of a quantum phase transition between a normal and a superfluid state~\cite{bruun14,bjerlin16,bayha20}. Indeed, mean-field theory predicts that the pair excitation becomes gapless at a critical coupling $G_{\rm crit}$~\cite{bruun14}, with a gapped excitation at \mbox{$\Delta E_{\rm ex} = 2 \Delta$} in the superfluid state that corresponds to an amplitude fluctuation of the order parameter, a Higgs excitation. We include this mean-field result in Fig.~\ref{fig:3}(a) as a gray dashed line, where $\Delta$ follows from Eq.~\eqref{eq:gapequation}. While mean-field theory should become reliable in the bulk limit of negligible level spacing \mbox{$\Delta\gg1$}~\cite{matveev97,bruun02,larkin05}, it already agrees qualitatively with our results at strong coupling, thus corroborating the interpretation of the pair excitation as a few-body Higgs precursor. Moreover, our results predict a higher pair excitation [green line in Fig.~\ref{fig:3}(a)] at almost twice the single-pair excitation energy, corresponding to a few-body precursor of a multiple Higgs excitation. Beyond these qualitative observations for small particle number, the pairing model~\eqref{eq:hamiltonian} now allows to investigate the full few-to-many-body crossover, which we describe in the following. 

Figure~\ref{fig:4} shows the evolution of the pair excitation for $15$ closed-shell configurations with \mbox{$N=6$}, $12,20,30,42,56,72,90,110,132,156,182,210,240$, and $272$ (gray lines).  
For larger particle numbers, the gap decreases, and the position of the minimum shifts to smaller coupling. Indeed, it extrapolates to infinite particle number in a power-law form \mbox{$E_{\rm min} \sim G_{\rm min}^\alpha$} with \mbox{$\alpha = 0.29(1)$} (bold blue line). This behavior is consistent with the expectation that in the thermodynamic limit, the pairing instability occurs at arbitrarily weak interaction strength~\cite{cooper56}. An important benchmark of our results are configuration-interaction calculations in the few-body limit~\cite{bjerlin16} shown as orange and purple crosses, where the latter calculations restrict the Hilbert space to three shells near the Fermi level. In both cases, we identify an effective interaction strength $G$ by relating the coupling in the respective calculations to the Fermi energy as \mbox{$G = A\varepsilon_b/n_F$} or \mbox{$G = A g/n_F$} with a joint fit parameter $A$ for different $N$, where $\varepsilon_b$ is the bound-state energy (in units of the trap frequency) and $g$ a bare contact interaction parameter. In all cases, the location of the minimum is quite well reproduced, indicating that the pairing model accurately captures pairing interactions at the Fermi level. Moreover, while there is a deviation of $10 \%$  for the minimum excitation energy between our calculations and the results of Ref.~\cite{bjerlin16} for \mbox{$N=6$} (indicating the relevance of pair-breaking matrix elements neglected in the pairing model), the agreement improves to $2\%$ with the three-shell calculation at \mbox{$N=30$}. This improved agreement with increasing $N$ suggests that the pairing model complements exact few-body calculations and provides a reliable extrapolation to large particle numbers. In addition, data points in the figure (green diamonds) show experimental measurements for \mbox{$N=6$} and \mbox{$N=12$} by~\textcite{bayha20}. We are able to fit the experimental data  by including a small trap anisotropy of $1\%$ for \mbox{$N=6$} [i.e., with trap frequencies that obey \mbox{$(\omega_x - \omega_y)/(\omega_x + \omega_y) = 1.01$}] and of \mbox{$0.4\%$} for \mbox{$N=12$} in our calculations (thin green lines), in line with the experimental anisotropy~\cite{bayha20}. In addition to this good agreement of our solution with the experimental result for the first excited state, we note that the Richardson model also captures several experimentally observed higher excitation branches: In particular, a higher non-monotonic branch observed in~\cite{bayha20}, which is interpreted as a pair excitation with finite angular momentum, corresponds to the second-lowest branch in Fig.~\ref{fig:3}(a) starting at excitation energy $2\hbar\omega$, which is derived from a parent state with a single pair in a higher level and two singly occupied valence states (the third from the left of the bottom inset states). In addition, Ref.~\cite{bayha20} also observes a higher branch with an excitation energy larger than twice the trap frequency, which is interpreted as a pair breaking excitation. Again, this excitation branch is present in our excitation spectrum in Fig.~\ref{fig:3}(a).

%++++++++++++++++++++++++++++++++++++++++
\begin{figure}[t!]
\includegraphics{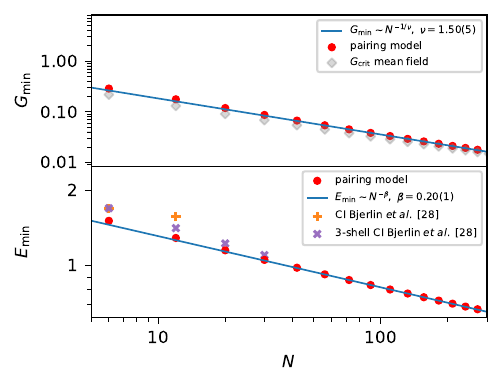}
\caption{(a) Position and (b) size of the minimum of the pair excitation energy as a function of particle number.}
\label{fig:5}
\end{figure}
%++++++++++++++++++++++++++++++++++++++++
 
To describe the few-to-many crossover quantitatively, we show in Fig.~\ref{fig:5}(a) the position of the minimum of the pair excitation in Fig.~\ref{fig:4} as a function of particle number, and Fig.~\ref{fig:5}(b) shows the corresponding minimum in the excitation energy. Both quantities vanish as a power law with increasing particle number: First, for the position of the minimum, we obtain \mbox{$G_{\rm min} \sim N^{-1/\nu}$} with \mbox{$\nu = 1.50(5)$} (blue line). The location of the minimum compares well with the mean-field prediction for the critical coupling (gray points), which further corroborates the interpretation of the pairing excitation as the few-body precursor of a quantum phase transition. Second, the excitation minimum decreases as \mbox{$E_{\rm min} \sim N^{-\beta}$} with an exponent \mbox{$\beta = 0.20(1)$} (blue line) that is characteristic of the finite-size scaling of the superfluid transition.

In summary, we provide a description of trapped Fermi gases with attractive pairing interactions that captures the full crossover between the few-body and the many-body regime. Our results indicate that the finite-size scaling of the quantum phase transition sets in early and is observable in experiments with few-body Fermi gases. Moreover, the pairing model used here is very versatile, in particular, it can be used to compute correlation functions~\cite{zhou02,faribault08} as recently measured in Ref.~\cite{holten22}.

\begin{acknowledgments}
We thank Johannes Bjerlin for discussions and Stephanie Reimann for discussions and comments on the manuscript. This work is supported by Vetenskapsr\aa det (Grant No. 2020-04239).
\end{acknowledgments}

\bibliography{bib_richardson}

\begin{thebibliography}{49}%
\makeatletter
\providecommand \@ifxundefined [1]{%
 \@ifx{#1\undefined}
}%
\providecommand \@ifnum [1]{%
 \ifnum #1\expandafter \@firstoftwo
 \else \expandafter \@secondoftwo
 \fi
}%
\providecommand \@ifx [1]{%
 \ifx #1\expandafter \@firstoftwo
 \else \expandafter \@secondoftwo
 \fi
}%
\providecommand \natexlab [1]{#1}%
\providecommand \enquote  [1]{``#1''}%
\providecommand \bibnamefont  [1]{#1}%
\providecommand \bibfnamefont [1]{#1}%
\providecommand \citenamefont [1]{#1}%
\providecommand \href@noop [0]{\@secondoftwo}%
\providecommand \href [0]{\begingroup \@sanitize@url \@href}%
\providecommand \@href[1]{\@@startlink{#1}\@@href}%
\providecommand \@@href[1]{\endgroup#1\@@endlink}%
\providecommand \@sanitize@url [0]{\catcode `\\12\catcode `\$12\catcode
  `\&12\catcode `\#12\catcode `\^12\catcode `\_12\catcode `\%12\relax}%
\providecommand \@@startlink[1]{}%
\providecommand \@@endlink[0]{}%
\providecommand \url  [0]{\begingroup\@sanitize@url \@url }%
\providecommand \@url [1]{\endgroup\@href {#1}{\urlprefix }}%
\providecommand \urlprefix  [0]{URL }%
\providecommand \Eprint [0]{\href }%
\providecommand \doibase [0]{https://doi.org/}%
\providecommand \selectlanguage [0]{\@gobble}%
\providecommand \bibinfo  [0]{\@secondoftwo}%
\providecommand \bibfield  [0]{\@secondoftwo}%
\providecommand \translation [1]{[#1]}%
\providecommand \BibitemOpen [0]{}%
\providecommand \bibitemStop [0]{}%
\providecommand \bibitemNoStop [0]{.\EOS\space}%
\providecommand \EOS [0]{\spacefactor3000\relax}%
\providecommand \BibitemShut  [1]{\csname bibitem#1\endcsname}%
\let\auto@bib@innerbib\@empty
%</preamble>
\bibitem [{\citenamefont {Anderson}(1972)}]{anderson72}%
  \BibitemOpen
  \bibfield  {author} {\bibinfo {author} {\bibfnamefont {P.~W.}\ \bibnamefont
  {Anderson}},\ }\bibfield  {title} {\bibinfo {title} {{More Is Different}},\
  }\href {https://doi.org/10.1126/science.177.4047.393} {\bibfield  {journal}
  {\bibinfo  {journal} {Science}\ }\textbf {\bibinfo {volume} {177}},\ \bibinfo
  {pages} {393} (\bibinfo {year} {1972})}\BibitemShut {NoStop}%
\bibitem [{\citenamefont {Rademann}\ \emph {et~al.}(1987)\citenamefont
  {Rademann}, \citenamefont {Kaiser}, \citenamefont {Even},\ and\ \citenamefont
  {Hensel}}]{rademann87}%
  \BibitemOpen
  \bibfield  {author} {\bibinfo {author} {\bibfnamefont {K.}~\bibnamefont
  {Rademann}}, \bibinfo {author} {\bibfnamefont {B.}~\bibnamefont {Kaiser}},
  \bibinfo {author} {\bibfnamefont {U.}~\bibnamefont {Even}},\ and\ \bibinfo
  {author} {\bibfnamefont {F.}~\bibnamefont {Hensel}},\ }\bibfield  {title}
  {\bibinfo {title} {{Size dependence of the gradual transition to metallic
  properties in isolated mercury clusters}},\ }\href
  {https://doi.org/10.1103/PhysRevLett.59.2319} {\bibfield  {journal} {\bibinfo
   {journal} {Phys. Rev. Lett.}\ }\textbf {\bibinfo {volume} {59}},\ \bibinfo
  {pages} {2319} (\bibinfo {year} {1987})}\BibitemShut {NoStop}%
\bibitem [{\citenamefont {Alivisatos}(1996)}]{alivisatos96}%
  \BibitemOpen
  \bibfield  {author} {\bibinfo {author} {\bibfnamefont {A.~P.}\ \bibnamefont
  {Alivisatos}},\ }\bibfield  {title} {\bibinfo {title} {{Semiconductor
  Clusters, Nanocrystals, and Quantum Dots}},\ }\href
  {https://doi.org/10.1126/science.271.5251.933} {\bibfield  {journal}
  {\bibinfo  {journal} {Science}\ }\textbf {\bibinfo {volume} {271}},\ \bibinfo
  {pages} {933} (\bibinfo {year} {1996})}\BibitemShut {NoStop}%
\bibitem [{\citenamefont {Reimann}\ and\ \citenamefont
  {Manninen}(2002)}]{reimann02}%
  \BibitemOpen
  \bibfield  {author} {\bibinfo {author} {\bibfnamefont {S.~M.}\ \bibnamefont
  {Reimann}}\ and\ \bibinfo {author} {\bibfnamefont {M.}~\bibnamefont
  {Manninen}},\ }\bibfield  {title} {\bibinfo {title} {{Electronic structure of
  quantum dots}},\ }\href {https://doi.org/10.1103/RevModPhys.74.1283}
  {\bibfield  {journal} {\bibinfo  {journal} {Rev. Mod. Phys.}\ }\textbf
  {\bibinfo {volume} {74}},\ \bibinfo {pages} {1283} (\bibinfo {year}
  {2002})}\BibitemShut {NoStop}%
\bibitem [{\citenamefont {Ralph}\ \emph {et~al.}(1995)\citenamefont {Ralph},
  \citenamefont {Black},\ and\ \citenamefont {Tinkham}}]{ralph95}%
  \BibitemOpen
  \bibfield  {author} {\bibinfo {author} {\bibfnamefont {D.~C.}\ \bibnamefont
  {Ralph}}, \bibinfo {author} {\bibfnamefont {C.~T.}\ \bibnamefont {Black}},\
  and\ \bibinfo {author} {\bibfnamefont {M.}~\bibnamefont {Tinkham}},\
  }\bibfield  {title} {\bibinfo {title} {{Spectroscopic Measurements of
  Discrete Electronic States in Single Metal Particles}},\ }\href
  {https://doi.org/10.1103/PhysRevLett.74.3241} {\bibfield  {journal} {\bibinfo
   {journal} {Phys. Rev. Lett.}\ }\textbf {\bibinfo {volume} {74}},\ \bibinfo
  {pages} {3241} (\bibinfo {year} {1995})}\BibitemShut {NoStop}%
\bibitem [{\citenamefont {Black}\ \emph {et~al.}(1996)\citenamefont {Black},
  \citenamefont {Ralph},\ and\ \citenamefont {Tinkham}}]{black96}%
  \BibitemOpen
  \bibfield  {author} {\bibinfo {author} {\bibfnamefont {C.~T.}\ \bibnamefont
  {Black}}, \bibinfo {author} {\bibfnamefont {D.~C.}\ \bibnamefont {Ralph}},\
  and\ \bibinfo {author} {\bibfnamefont {M.}~\bibnamefont {Tinkham}},\
  }\bibfield  {title} {\bibinfo {title} {{Spectroscopy of the Superconducting
  Gap in Individual Nanometer-Scale Aluminum Particles}},\ }\href
  {https://doi.org/10.1103/PhysRevLett.76.688} {\bibfield  {journal} {\bibinfo
  {journal} {Phys. Rev. Lett.}\ }\textbf {\bibinfo {volume} {76}},\ \bibinfo
  {pages} {688} (\bibinfo {year} {1996})}\BibitemShut {NoStop}%
\bibitem [{\citenamefont {Ralph}\ \emph {et~al.}(1997)\citenamefont {Ralph},
  \citenamefont {Black},\ and\ \citenamefont {Tinkham}}]{ralph97}%
  \BibitemOpen
  \bibfield  {author} {\bibinfo {author} {\bibfnamefont {D.~C.}\ \bibnamefont
  {Ralph}}, \bibinfo {author} {\bibfnamefont {C.~T.}\ \bibnamefont {Black}},\
  and\ \bibinfo {author} {\bibfnamefont {M.}~\bibnamefont {Tinkham}},\
  }\bibfield  {title} {\bibinfo {title} {{Gate-Voltage Studies of Discrete
  Electronic States in Aluminum Nanoparticles}},\ }\href
  {https://doi.org/10.1103/PhysRevLett.78.4087} {\bibfield  {journal} {\bibinfo
   {journal} {Phys. Rev. Lett.}\ }\textbf {\bibinfo {volume} {78}},\ \bibinfo
  {pages} {4087} (\bibinfo {year} {1997})}\BibitemShut {NoStop}%
\bibitem [{\citenamefont {Serwane}\ \emph {et~al.}(2011)\citenamefont
  {Serwane}, \citenamefont {Z{\"u}rn}, \citenamefont {Lompe}, \citenamefont
  {Ottenstein}, \citenamefont {Wenz},\ and\ \citenamefont
  {Jochim}}]{serwane11}%
  \BibitemOpen
  \bibfield  {author} {\bibinfo {author} {\bibfnamefont {F.}~\bibnamefont
  {Serwane}}, \bibinfo {author} {\bibfnamefont {G.}~\bibnamefont {Z{\"u}rn}},
  \bibinfo {author} {\bibfnamefont {T.}~\bibnamefont {Lompe}}, \bibinfo
  {author} {\bibfnamefont {T.~B.}\ \bibnamefont {Ottenstein}}, \bibinfo
  {author} {\bibfnamefont {A.~N.}\ \bibnamefont {Wenz}},\ and\ \bibinfo
  {author} {\bibfnamefont {S.}~\bibnamefont {Jochim}},\ }\bibfield  {title}
  {\bibinfo {title} {{Deterministic Preparation of a Tunable Few-Fermion
  System}},\ }\href {https://doi.org/10.1126/science.1201351} {\bibfield
  {journal} {\bibinfo  {journal} {Science}\ }\textbf {\bibinfo {volume}
  {332}},\ \bibinfo {pages} {336} (\bibinfo {year} {2011})}\BibitemShut
  {NoStop}%
\bibitem [{\citenamefont {Z\"urn}\ \emph {et~al.}(2013)\citenamefont {Z\"urn},
  \citenamefont {Wenz}, \citenamefont {Murmann}, \citenamefont {Bergschneider},
  \citenamefont {Lompe},\ and\ \citenamefont {Jochim}}]{zuern13}%
  \BibitemOpen
  \bibfield  {author} {\bibinfo {author} {\bibfnamefont {G.}~\bibnamefont
  {Z\"urn}}, \bibinfo {author} {\bibfnamefont {A.~N.}\ \bibnamefont {Wenz}},
  \bibinfo {author} {\bibfnamefont {S.}~\bibnamefont {Murmann}}, \bibinfo
  {author} {\bibfnamefont {A.}~\bibnamefont {Bergschneider}}, \bibinfo {author}
  {\bibfnamefont {T.}~\bibnamefont {Lompe}},\ and\ \bibinfo {author}
  {\bibfnamefont {S.}~\bibnamefont {Jochim}},\ }\bibfield  {title} {\bibinfo
  {title} {{Pairing in Few-Fermion Systems with Attractive Interactions}},\
  }\href {https://doi.org/10.1103/PhysRevLett.111.175302} {\bibfield  {journal}
  {\bibinfo  {journal} {Phys. Rev. Lett.}\ }\textbf {\bibinfo {volume} {111}},\
  \bibinfo {pages} {175302} (\bibinfo {year} {2013})}\BibitemShut {NoStop}%
\bibitem [{\citenamefont {Bayha}\ \emph {et~al.}(2020)\citenamefont {Bayha},
  \citenamefont {Holten}, \citenamefont {Klemt}, \citenamefont {Subramanian},
  \citenamefont {Bjerlin}, \citenamefont {Reimann}, \citenamefont {Bruun},
  \citenamefont {Preiss},\ and\ \citenamefont {Jochim}}]{bayha20}%
  \BibitemOpen
  \bibfield  {author} {\bibinfo {author} {\bibfnamefont {L.}~\bibnamefont
  {Bayha}}, \bibinfo {author} {\bibfnamefont {M.}~\bibnamefont {Holten}},
  \bibinfo {author} {\bibfnamefont {R.}~\bibnamefont {Klemt}}, \bibinfo
  {author} {\bibfnamefont {K.}~\bibnamefont {Subramanian}}, \bibinfo {author}
  {\bibfnamefont {J.}~\bibnamefont {Bjerlin}}, \bibinfo {author} {\bibfnamefont
  {S.~M.}\ \bibnamefont {Reimann}}, \bibinfo {author} {\bibfnamefont {G.~M.}\
  \bibnamefont {Bruun}}, \bibinfo {author} {\bibfnamefont {P.~M.}\ \bibnamefont
  {Preiss}},\ and\ \bibinfo {author} {\bibfnamefont {S.}~\bibnamefont
  {Jochim}},\ }\bibfield  {title} {\bibinfo {title} {{Observing the emergence
  of a quantum phase transition shell by shell}},\ }\href
  {https://doi.org/10.1038/s41586-020-2936-y} {\bibfield  {journal} {\bibinfo
  {journal} {Nature (London)}\ }\textbf {\bibinfo {volume} {587}},\ \bibinfo
  {pages} {583} (\bibinfo {year} {2020})}\BibitemShut {NoStop}%
\bibitem [{\citenamefont {Holten}\ \emph {et~al.}(2021)\citenamefont {Holten},
  \citenamefont {Bayha}, \citenamefont {Subramanian}, \citenamefont {Heintze},
  \citenamefont {Preiss},\ and\ \citenamefont {Jochim}}]{holten21a}%
  \BibitemOpen
  \bibfield  {author} {\bibinfo {author} {\bibfnamefont {M.}~\bibnamefont
  {Holten}}, \bibinfo {author} {\bibfnamefont {L.}~\bibnamefont {Bayha}},
  \bibinfo {author} {\bibfnamefont {K.}~\bibnamefont {Subramanian}}, \bibinfo
  {author} {\bibfnamefont {C.}~\bibnamefont {Heintze}}, \bibinfo {author}
  {\bibfnamefont {P.~M.}\ \bibnamefont {Preiss}},\ and\ \bibinfo {author}
  {\bibfnamefont {S.}~\bibnamefont {Jochim}},\ }\bibfield  {title} {\bibinfo
  {title} {{Observation of Pauli Crystals}},\ }\href
  {https://doi.org/10.1103/PhysRevLett.126.020401} {\bibfield  {journal}
  {\bibinfo  {journal} {Phys. Rev. Lett.}\ }\textbf {\bibinfo {volume} {126}},\
  \bibinfo {pages} {020401} (\bibinfo {year} {2021})}\BibitemShut {NoStop}%
\bibitem [{\citenamefont {Holten}\ \emph {et~al.}(2022)\citenamefont {Holten},
  \citenamefont {Bayha}, \citenamefont {Subramanian}, \citenamefont
  {Brandstetter}, \citenamefont {Heintze}, \citenamefont {Lunt}, \citenamefont
  {Preiss},\ and\ \citenamefont {Jochim}}]{holten22}%
  \BibitemOpen
  \bibfield  {author} {\bibinfo {author} {\bibfnamefont {M.}~\bibnamefont
  {Holten}}, \bibinfo {author} {\bibfnamefont {L.}~\bibnamefont {Bayha}},
  \bibinfo {author} {\bibfnamefont {K.}~\bibnamefont {Subramanian}}, \bibinfo
  {author} {\bibfnamefont {S.}~\bibnamefont {Brandstetter}}, \bibinfo {author}
  {\bibfnamefont {C.}~\bibnamefont {Heintze}}, \bibinfo {author} {\bibfnamefont
  {P.}~\bibnamefont {Lunt}}, \bibinfo {author} {\bibfnamefont {P.~M.}\
  \bibnamefont {Preiss}},\ and\ \bibinfo {author} {\bibfnamefont
  {S.}~\bibnamefont {Jochim}},\ }\bibfield  {title} {\bibinfo {title}
  {{Observation of Cooper pairs in a mesoscopic two-dimensional Fermi gas}},\
  }\href {https://doi.org/10.1038/s41586-022-04678-1} {\bibfield  {journal}
  {\bibinfo  {journal} {Nature}\ }\textbf {\bibinfo {volume} {606}},\ \bibinfo
  {pages} {287} (\bibinfo {year} {2022})}\BibitemShut {NoStop}%
\bibitem [{\citenamefont {Bruun}(2014)}]{bruun14}%
  \BibitemOpen
  \bibfield  {author} {\bibinfo {author} {\bibfnamefont {G.~M.}\ \bibnamefont
  {Bruun}},\ }\bibfield  {title} {\bibinfo {title} {{Long-lived Higgs mode in a
  two-dimensional confined Fermi system}},\ }\href
  {https://doi.org/10.1103/PhysRevA.90.023621} {\bibfield  {journal} {\bibinfo
  {journal} {Phys. Rev. A}\ }\textbf {\bibinfo {volume} {90}},\ \bibinfo
  {pages} {023621} (\bibinfo {year} {2014})}\BibitemShut {NoStop}%
\bibitem [{\citenamefont {Hofmann}\ \emph {et~al.}(2016)\citenamefont
  {Hofmann}, \citenamefont {Lobos},\ and\ \citenamefont
  {Galitski}}]{hofmann16}%
  \BibitemOpen
  \bibfield  {author} {\bibinfo {author} {\bibfnamefont {J.}~\bibnamefont
  {Hofmann}}, \bibinfo {author} {\bibfnamefont {A.~M.}\ \bibnamefont {Lobos}},\
  and\ \bibinfo {author} {\bibfnamefont {V.}~\bibnamefont {Galitski}},\
  }\bibfield  {title} {\bibinfo {title} {{Parity effect in a mesoscopic Fermi
  gas}},\ }\href {https://doi.org/10.1103/PhysRevA.93.061602} {\bibfield
  {journal} {\bibinfo  {journal} {Phys. Rev. A}\ }\textbf {\bibinfo {volume}
  {93}},\ \bibinfo {pages} {061602} (\bibinfo {year} {2016})}\BibitemShut
  {NoStop}%
\bibitem [{\citenamefont {Hofmann}(2017)}]{hofmann17}%
  \BibitemOpen
  \bibfield  {author} {\bibinfo {author} {\bibfnamefont {J.}~\bibnamefont
  {Hofmann}},\ }\bibfield  {title} {\bibinfo {title} {{Mesoscopic pairing
  without superconductivity}},\ }\href
  {https://doi.org/10.1103/PhysRevB.96.220508} {\bibfield  {journal} {\bibinfo
  {journal} {Phys. Rev. B}\ }\textbf {\bibinfo {volume} {96}},\ \bibinfo
  {pages} {220508} (\bibinfo {year} {2017})}\BibitemShut {NoStop}%
\bibitem [{\citenamefont {Richardson}(1963)}]{richardson63}%
  \BibitemOpen
  \bibfield  {author} {\bibinfo {author} {\bibfnamefont {R.~W.}\ \bibnamefont
  {Richardson}},\ }\bibfield  {title} {\bibinfo {title} {{A Restricted Class of
  Exact Eigenstates of the Pairing-Force Hamiltonian}},\ }\href
  {http://www.sciencedirect.com/science/article/pii/0031916363902592}
  {\bibfield  {journal} {\bibinfo  {journal} {Phys. Lett.}\ }\textbf {\bibinfo
  {volume} {3}},\ \bibinfo {pages} {277} (\bibinfo {year} {1963})}\BibitemShut
  {NoStop}%
\bibitem [{\citenamefont {Richardson}\ and\ \citenamefont
  {Sherman}(1964)}]{richardson64}%
  \BibitemOpen
  \bibfield  {author} {\bibinfo {author} {\bibfnamefont {R.~W.}\ \bibnamefont
  {Richardson}}\ and\ \bibinfo {author} {\bibfnamefont {N.}~\bibnamefont
  {Sherman}},\ }\bibfield  {title} {\bibinfo {title} {{Exact Eigenstates of the
  Pairing-Force Hamiltonian}},\ }\href
  {http://www.sciencedirect.com/science/article/pii/002955826490687X}
  {\bibfield  {journal} {\bibinfo  {journal} {Nucl. Phys.}\ }\textbf {\bibinfo
  {volume} {52}},\ \bibinfo {pages} {221} (\bibinfo {year} {1964})}\BibitemShut
  {NoStop}%
\bibitem [{\citenamefont {Richardson}(1965)}]{richardson65}%
  \BibitemOpen
  \bibfield  {author} {\bibinfo {author} {\bibfnamefont {R.~W.}\ \bibnamefont
  {Richardson}},\ }\bibfield  {title} {\bibinfo {title} {{Exact Eigenstates of
  the Pairing Force Hamiltonian. II}},\ }\href
  {http://dx.doi.org/10.1063/1.1704367} {\bibfield  {journal} {\bibinfo
  {journal} {J. Math. Phys.}\ }\textbf {\bibinfo {volume} {6}},\ \bibinfo
  {pages} {1034} (\bibinfo {year} {1965})}\BibitemShut {NoStop}%
\bibitem [{\citenamefont {Richardson}(1966)}]{richardson66}%
  \BibitemOpen
  \bibfield  {author} {\bibinfo {author} {\bibfnamefont {R.~W.}\ \bibnamefont
  {Richardson}},\ }\bibfield  {title} {\bibinfo {title} {{Numerical Study of
  the 8-32-Particle Eigenstates of the Pairing Hamiltonian}},\ }\href
  {http://link.aps.org/doi/10.1103/PhysRev.141.949} {\bibfield  {journal}
  {\bibinfo  {journal} {Phys. Rev.}\ }\textbf {\bibinfo {volume} {141}},\
  \bibinfo {pages} {949} (\bibinfo {year} {1966})}\BibitemShut {NoStop}%
\bibitem [{\citenamefont {{von Delft}}\ and\ \citenamefont
  {Ralph}(2001)}]{vondelft01}%
  \BibitemOpen
  \bibfield  {author} {\bibinfo {author} {\bibfnamefont {J.}~\bibnamefont {{von
  Delft}}}\ and\ \bibinfo {author} {\bibfnamefont {D.~C.}\ \bibnamefont
  {Ralph}},\ }\bibfield  {title} {\bibinfo {title} {{Spectroscopy of discrete
  energy levels in ultrasmall metallic grains}},\ }\href
  {https://doi.org/https://doi.org/10.1016/S0370-1573(00)00099-5} {\bibfield
  {journal} {\bibinfo  {journal} {Physics Reports}\ }\textbf {\bibinfo {volume}
  {345}},\ \bibinfo {pages} {61} (\bibinfo {year} {2001})}\BibitemShut
  {NoStop}%
\bibitem [{\citenamefont {Dukelsky}\ \emph {et~al.}(2004)\citenamefont
  {Dukelsky}, \citenamefont {Pittel},\ and\ \citenamefont
  {Sierra}}]{dukelsky04}%
  \BibitemOpen
  \bibfield  {author} {\bibinfo {author} {\bibfnamefont {J.}~\bibnamefont
  {Dukelsky}}, \bibinfo {author} {\bibfnamefont {S.}~\bibnamefont {Pittel}},\
  and\ \bibinfo {author} {\bibfnamefont {G.}~\bibnamefont {Sierra}},\
  }\bibfield  {title} {\bibinfo {title} {{Colloquium: Exactly solvable
  Richardson-Gaudin models for many-body quantum systems}},\ }\href
  {https://doi.org/10.1103/RevModPhys.76.643} {\bibfield  {journal} {\bibinfo
  {journal} {Rev. Mod. Phys.}\ }\textbf {\bibinfo {volume} {76}},\ \bibinfo
  {pages} {643} (\bibinfo {year} {2004})}\BibitemShut {NoStop}%
\bibitem [{\citenamefont {Johnson}(2023)}]{johnson23}%
  \BibitemOpen
  \bibfield  {author} {\bibinfo {author} {\bibfnamefont {P.~A.}\ \bibnamefont
  {Johnson}},\ }\bibfield  {title} {\bibinfo {title} {{Richardson-Gaudin
  States}},\ }\bibfield  {journal} {\bibinfo  {journal} {arXiv preprint
  arXiv:2312.08804}\ }\href {https://doi.org/10.48550/arXiv.2312.08804}
  {10.48550/arXiv.2312.08804} (\bibinfo {year} {2023})\BibitemShut {NoStop}%
\bibitem [{\citenamefont {Rontani}\ \emph {et~al.}(2009)\citenamefont
  {Rontani}, \citenamefont {Armstrong}, \citenamefont {Yu}, \citenamefont
  {\AA{}berg},\ and\ \citenamefont {Reimann}}]{rontani09}%
  \BibitemOpen
  \bibfield  {author} {\bibinfo {author} {\bibfnamefont {M.}~\bibnamefont
  {Rontani}}, \bibinfo {author} {\bibfnamefont {J.~R.}\ \bibnamefont
  {Armstrong}}, \bibinfo {author} {\bibfnamefont {Y.}~\bibnamefont {Yu}},
  \bibinfo {author} {\bibfnamefont {S.}~\bibnamefont {\AA{}berg}},\ and\
  \bibinfo {author} {\bibfnamefont {S.~M.}\ \bibnamefont {Reimann}},\
  }\bibfield  {title} {\bibinfo {title} {{Cold Fermionic Atoms in
  Two-Dimensional Traps: Pairing versus Hund's Rule}},\ }\href
  {https://doi.org/10.1103/PhysRevLett.102.060401} {\bibfield  {journal}
  {\bibinfo  {journal} {Phys. Rev. Lett.}\ }\textbf {\bibinfo {volume} {102}},\
  \bibinfo {pages} {060401} (\bibinfo {year} {2009})}\BibitemShut {NoStop}%
\bibitem [{\citenamefont {Sowi\ifmmode~\acute{n}\else \'{n}\fi{}ski}\ \emph
  {et~al.}(2013)\citenamefont {Sowi\ifmmode~\acute{n}\else \'{n}\fi{}ski},
  \citenamefont {Grass}, \citenamefont {Dutta},\ and\ \citenamefont
  {Lewenstein}}]{sowinski13}%
  \BibitemOpen
  \bibfield  {author} {\bibinfo {author} {\bibfnamefont {T.}~\bibnamefont
  {Sowi\ifmmode~\acute{n}\else \'{n}\fi{}ski}}, \bibinfo {author}
  {\bibfnamefont {T.}~\bibnamefont {Grass}}, \bibinfo {author} {\bibfnamefont
  {O.}~\bibnamefont {Dutta}},\ and\ \bibinfo {author} {\bibfnamefont
  {M.}~\bibnamefont {Lewenstein}},\ }\bibfield  {title} {\bibinfo {title} {{Few
  interacting fermions in a one-dimensional harmonic trap}},\ }\href
  {https://doi.org/10.1103/PhysRevA.88.033607} {\bibfield  {journal} {\bibinfo
  {journal} {Phys. Rev. A}\ }\textbf {\bibinfo {volume} {88}},\ \bibinfo
  {pages} {033607} (\bibinfo {year} {2013})}\BibitemShut {NoStop}%
\bibitem [{\citenamefont {D'Amico}\ and\ \citenamefont
  {Rontani}(2015)}]{damico15}%
  \BibitemOpen
  \bibfield  {author} {\bibinfo {author} {\bibfnamefont {P.}~\bibnamefont
  {D'Amico}}\ and\ \bibinfo {author} {\bibfnamefont {M.}~\bibnamefont
  {Rontani}},\ }\bibfield  {title} {\bibinfo {title} {{Pairing of a few Fermi
  atoms in one dimension}},\ }\href
  {https://doi.org/10.1103/PhysRevA.91.043610} {\bibfield  {journal} {\bibinfo
  {journal} {Phys. Rev. A}\ }\textbf {\bibinfo {volume} {91}},\ \bibinfo
  {pages} {043610} (\bibinfo {year} {2015})}\BibitemShut {NoStop}%
\bibitem [{\citenamefont {Sowi{\'{n}}ski}\ \emph {et~al.}(2015)\citenamefont
  {Sowi{\'{n}}ski}, \citenamefont {Gajda},\ and\ \citenamefont
  {Rza{\.{z}}ewski}}]{sowinski15}%
  \BibitemOpen
  \bibfield  {author} {\bibinfo {author} {\bibfnamefont {T.}~\bibnamefont
  {Sowi{\'{n}}ski}}, \bibinfo {author} {\bibfnamefont {M.}~\bibnamefont
  {Gajda}},\ and\ \bibinfo {author} {\bibfnamefont {K.}~\bibnamefont
  {Rza{\.{z}}ewski}},\ }\bibfield  {title} {\bibinfo {title} {{Pairing in a
  system of a few attractive fermions in a harmonic trap}},\ }\href
  {https://doi.org/10.1209/0295-5075/109/26005} {\bibfield  {journal} {\bibinfo
   {journal} {{EPL} (Europhysics Letters)}\ }\textbf {\bibinfo {volume}
  {109}},\ \bibinfo {pages} {26005} (\bibinfo {year} {2015})}\BibitemShut
  {NoStop}%
\bibitem [{\citenamefont {Grining}\ \emph {et~al.}(2015)\citenamefont
  {Grining}, \citenamefont {Tomza}, \citenamefont {Lesiuk}, \citenamefont
  {Przybytek}, \citenamefont {Musia\l{}}, \citenamefont {Moszynski},
  \citenamefont {Lewenstein},\ and\ \citenamefont {Massignan}}]{grining15}%
  \BibitemOpen
  \bibfield  {author} {\bibinfo {author} {\bibfnamefont {T.}~\bibnamefont
  {Grining}}, \bibinfo {author} {\bibfnamefont {M.}~\bibnamefont {Tomza}},
  \bibinfo {author} {\bibfnamefont {M.}~\bibnamefont {Lesiuk}}, \bibinfo
  {author} {\bibfnamefont {M.}~\bibnamefont {Przybytek}}, \bibinfo {author}
  {\bibfnamefont {M.}~\bibnamefont {Musia\l{}}}, \bibinfo {author}
  {\bibfnamefont {R.}~\bibnamefont {Moszynski}}, \bibinfo {author}
  {\bibfnamefont {M.}~\bibnamefont {Lewenstein}},\ and\ \bibinfo {author}
  {\bibfnamefont {P.}~\bibnamefont {Massignan}},\ }\bibfield  {title} {\bibinfo
  {title} {{Crossover between few and many fermions in a harmonic trap}},\
  }\href {https://doi.org/10.1103/PhysRevA.92.061601} {\bibfield  {journal}
  {\bibinfo  {journal} {Phys. Rev. A}\ }\textbf {\bibinfo {volume} {92}},\
  \bibinfo {pages} {061601} (\bibinfo {year} {2015})}\BibitemShut {NoStop}%
\bibitem [{\citenamefont {Bjerlin}\ \emph {et~al.}(2016)\citenamefont
  {Bjerlin}, \citenamefont {Reimann},\ and\ \citenamefont {Bruun}}]{bjerlin16}%
  \BibitemOpen
  \bibfield  {author} {\bibinfo {author} {\bibfnamefont {J.}~\bibnamefont
  {Bjerlin}}, \bibinfo {author} {\bibfnamefont {S.~M.}\ \bibnamefont
  {Reimann}},\ and\ \bibinfo {author} {\bibfnamefont {G.~M.}\ \bibnamefont
  {Bruun}},\ }\bibfield  {title} {\bibinfo {title} {{Few-Body Precursor of the
  Higgs Mode in a Fermi Gas}},\ }\href
  {https://doi.org/10.1103/PhysRevLett.116.155302} {\bibfield  {journal}
  {\bibinfo  {journal} {Phys. Rev. Lett.}\ }\textbf {\bibinfo {volume} {116}},\
  \bibinfo {pages} {155302} (\bibinfo {year} {2016})}\BibitemShut {NoStop}%
\bibitem [{\citenamefont {Rontani}\ \emph {et~al.}(2017)\citenamefont
  {Rontani}, \citenamefont {Eriksson}, \citenamefont {{\AA}berg},\ and\
  \citenamefont {Reimann}}]{rontani17}%
  \BibitemOpen
  \bibfield  {author} {\bibinfo {author} {\bibfnamefont {M.}~\bibnamefont
  {Rontani}}, \bibinfo {author} {\bibfnamefont {G.}~\bibnamefont {Eriksson}},
  \bibinfo {author} {\bibfnamefont {S.}~\bibnamefont {{\AA}berg}},\ and\
  \bibinfo {author} {\bibfnamefont {S.~M.}\ \bibnamefont {Reimann}},\
  }\bibfield  {title} {\bibinfo {title} {{On the renormalization of contact
  interactions for the configuration-interaction method in two-dimensions}},\
  }\href {https://doi.org/10.1088/1361-6455/aa606a} {\bibfield  {journal}
  {\bibinfo  {journal} {Journal of Physics B: Atomic, Molecular and Optical
  Physics}\ }\textbf {\bibinfo {volume} {50}},\ \bibinfo {pages} {065301}
  (\bibinfo {year} {2017})}\BibitemShut {NoStop}%
\bibitem [{\citenamefont {Bekassy}\ and\ \citenamefont
  {Hofmann}(2022)}]{bekassy22}%
  \BibitemOpen
  \bibfield  {author} {\bibinfo {author} {\bibfnamefont {V.}~\bibnamefont
  {Bekassy}}\ and\ \bibinfo {author} {\bibfnamefont {J.}~\bibnamefont
  {Hofmann}},\ }\bibfield  {title} {\bibinfo {title} {{Nonrelativistic
  Conformal Invariance in Mesoscopic Two-Dimensional Fermi Gases}},\ }\href
  {https://doi.org/10.1103/PhysRevLett.128.193401} {\bibfield  {journal}
  {\bibinfo  {journal} {Phys. Rev. Lett.}\ }\textbf {\bibinfo {volume} {128}},\
  \bibinfo {pages} {193401} (\bibinfo {year} {2022})}\BibitemShut {NoStop}%
\bibitem [{\citenamefont {Bekassy}\ and\ \citenamefont
  {Hofmann}(2024)}]{bekassy24}%
  \BibitemOpen
  \bibfield  {author} {\bibinfo {author} {\bibfnamefont {V.}~\bibnamefont
  {Bekassy}}\ and\ \bibinfo {author} {\bibfnamefont {J.}~\bibnamefont
  {Hofmann}},\ }\bibfield  {title} {\bibinfo {title} {{Scale and conformal
  invariance in rotating interacting few-fermion systems}},\ }\href
  {https://doi.org/10.1103/PhysRevResearch.6.023279} {\bibfield  {journal}
  {\bibinfo  {journal} {Phys. Rev. Res.}\ }\textbf {\bibinfo {volume} {6}},\
  \bibinfo {pages} {023279} (\bibinfo {year} {2024})}\BibitemShut {NoStop}%
\bibitem [{\citenamefont {Laird}\ \emph {et~al.}(2024)\citenamefont {Laird},
  \citenamefont {Mulkerin}, \citenamefont {Wang},\ and\ \citenamefont
  {Davis}}]{laird24}%
  \BibitemOpen
  \bibfield  {author} {\bibinfo {author} {\bibfnamefont {E.}~\bibnamefont
  {Laird}}, \bibinfo {author} {\bibfnamefont {B.}~\bibnamefont {Mulkerin}},
  \bibinfo {author} {\bibfnamefont {J.}~\bibnamefont {Wang}},\ and\ \bibinfo
  {author} {\bibfnamefont {M.}~\bibnamefont {Davis}},\ }\bibfield  {title}
  {\bibinfo {title} {{When does a Fermi puddle become a Fermi sea? Emergence of
  Pairing in Two-Dimensional Trapped Mesoscopic Fermi Gases}},\ }\bibfield
  {journal} {\bibinfo  {journal} {arXiv preprint arXiv:2408.17015}\ }\href
  {https://doi.org/10.48550/arXiv.2408.17015} {10.48550/arXiv.2408.17015}
  (\bibinfo {year} {2024})\BibitemShut {NoStop}%
\bibitem [{\citenamefont {Berger}\ \emph {et~al.}(2015)\citenamefont {Berger},
  \citenamefont {Anderson},\ and\ \citenamefont {Drut}}]{berger15}%
  \BibitemOpen
  \bibfield  {author} {\bibinfo {author} {\bibfnamefont {C.~E.}\ \bibnamefont
  {Berger}}, \bibinfo {author} {\bibfnamefont {E.~R.}\ \bibnamefont
  {Anderson}},\ and\ \bibinfo {author} {\bibfnamefont {J.~E.}\ \bibnamefont
  {Drut}},\ }\bibfield  {title} {\bibinfo {title} {{Energy, contact, and
  density profiles of one-dimensional fermions in a harmonic trap via
  nonuniform-lattice Monte Carlo calculations}},\ }\href
  {https://doi.org/10.1103/PhysRevA.91.053618} {\bibfield  {journal} {\bibinfo
  {journal} {Phys. Rev. A}\ }\textbf {\bibinfo {volume} {91}},\ \bibinfo
  {pages} {053618} (\bibinfo {year} {2015})}\BibitemShut {NoStop}%
\bibitem [{\citenamefont {Rammelm\"uller}\ \emph {et~al.}(2016)\citenamefont
  {Rammelm\"uller}, \citenamefont {Porter},\ and\ \citenamefont
  {Drut}}]{rammelmueller16}%
  \BibitemOpen
  \bibfield  {author} {\bibinfo {author} {\bibfnamefont {L.}~\bibnamefont
  {Rammelm\"uller}}, \bibinfo {author} {\bibfnamefont {W.~J.}\ \bibnamefont
  {Porter}},\ and\ \bibinfo {author} {\bibfnamefont {J.~E.}\ \bibnamefont
  {Drut}},\ }\bibfield  {title} {\bibinfo {title} {{Ground state of the
  two-dimensional attractive Fermi gas: Essential properties from few to many
  body}},\ }\href {https://doi.org/10.1103/PhysRevA.93.033639} {\bibfield
  {journal} {\bibinfo  {journal} {Phys. Rev. A}\ }\textbf {\bibinfo {volume}
  {93}},\ \bibinfo {pages} {033639} (\bibinfo {year} {2016})}\BibitemShut
  {NoStop}%
\bibitem [{\citenamefont {Luo}\ \emph {et~al.}(2016)\citenamefont {Luo},
  \citenamefont {Berger},\ and\ \citenamefont {Drut}}]{luo16}%
  \BibitemOpen
  \bibfield  {author} {\bibinfo {author} {\bibfnamefont {Z.}~\bibnamefont
  {Luo}}, \bibinfo {author} {\bibfnamefont {C.~E.}\ \bibnamefont {Berger}},\
  and\ \bibinfo {author} {\bibfnamefont {J.~E.}\ \bibnamefont {Drut}},\
  }\bibfield  {title} {\bibinfo {title} {{Harmonically trapped fermions in two
  dimensions: Ground-state energy and contact of SU(2) and SU(4) systems via a
  nonuniform lattice Monte Carlo method}},\ }\href
  {https://doi.org/10.1103/PhysRevA.93.033604} {\bibfield  {journal} {\bibinfo
  {journal} {Phys. Rev. A}\ }\textbf {\bibinfo {volume} {93}},\ \bibinfo
  {pages} {033604} (\bibinfo {year} {2016})}\BibitemShut {NoStop}%
\bibitem [{\citenamefont {Matveev}\ and\ \citenamefont
  {Larkin}(1997)}]{matveev97}%
  \BibitemOpen
  \bibfield  {author} {\bibinfo {author} {\bibfnamefont {K.~A.}\ \bibnamefont
  {Matveev}}\ and\ \bibinfo {author} {\bibfnamefont {A.~I.}\ \bibnamefont
  {Larkin}},\ }\bibfield  {title} {\bibinfo {title} {{Parity Effect in Ground
  State Energies of Ultrasmall Superconducting Grains}},\ }\href
  {https://doi.org/10.1103/PhysRevLett.78.3749} {\bibfield  {journal} {\bibinfo
   {journal} {Phys. Rev. Lett.}\ }\textbf {\bibinfo {volume} {78}},\ \bibinfo
  {pages} {3749} (\bibinfo {year} {1997})}\BibitemShut {NoStop}%
\bibitem [{\citenamefont {von Delft}\ and\ \citenamefont
  {Braun}(2000)}]{vondelft00}%
  \BibitemOpen
  \bibfield  {author} {\bibinfo {author} {\bibfnamefont {J.}~\bibnamefont {von
  Delft}}\ and\ \bibinfo {author} {\bibfnamefont {F.}~\bibnamefont {Braun}},\
  }\bibfield  {title} {\bibinfo {title} {{Superconductivity in Ultrasmall
  Grains: Introduction to Richardson's Exact Solution}},\ }in\ \href@noop {}
  {\emph {\bibinfo {booktitle} {Proceedings of the NATO ASI "Quantum Mesoscopic
  Phenomena and Mesoscopic Devices in Microelectronics"}}},\ \bibinfo {editor}
  {edited by\ \bibinfo {editor} {\bibfnamefont {I.~O.}\ \bibnamefont {Kulik}}\
  and\ \bibinfo {editor} {\bibfnamefont {R.}~\bibnamefont {Ellialtioglu}}}\
  (\bibinfo {organization} {Kluwer, Dordrecht},\ \bibinfo {year} {2000})\ p.\
  \bibinfo {pages} {361}\BibitemShut {NoStop}%
\bibitem [{\citenamefont {Faribault}\ \emph {et~al.}(2011)\citenamefont
  {Faribault}, \citenamefont {El~Araby}, \citenamefont {Str\"ater},\ and\
  \citenamefont {Gritsev}}]{faribault11}%
  \BibitemOpen
  \bibfield  {author} {\bibinfo {author} {\bibfnamefont {A.}~\bibnamefont
  {Faribault}}, \bibinfo {author} {\bibfnamefont {O.}~\bibnamefont {El~Araby}},
  \bibinfo {author} {\bibfnamefont {C.}~\bibnamefont {Str\"ater}},\ and\
  \bibinfo {author} {\bibfnamefont {V.}~\bibnamefont {Gritsev}},\ }\bibfield
  {title} {\bibinfo {title} {{Gaudin models solver based on the correspondence
  between Bethe ansatz and ordinary differential equations}},\ }\href
  {https://doi.org/10.1103/PhysRevB.83.235124} {\bibfield  {journal} {\bibinfo
  {journal} {Phys. Rev. B}\ }\textbf {\bibinfo {volume} {83}},\ \bibinfo
  {pages} {235124} (\bibinfo {year} {2011})}\BibitemShut {NoStop}%
\bibitem [{\citenamefont {El~Araby}\ \emph {et~al.}(2012)\citenamefont
  {El~Araby}, \citenamefont {Gritsev},\ and\ \citenamefont
  {Faribault}}]{elaraby12}%
  \BibitemOpen
  \bibfield  {author} {\bibinfo {author} {\bibfnamefont {O.}~\bibnamefont
  {El~Araby}}, \bibinfo {author} {\bibfnamefont {V.}~\bibnamefont {Gritsev}},\
  and\ \bibinfo {author} {\bibfnamefont {A.}~\bibnamefont {Faribault}},\
  }\bibfield  {title} {\bibinfo {title} {{Bethe ansatz and ordinary
  differential equation correspondence for degenerate Gaudin models}},\ }\href
  {https://doi.org/10.1103/PhysRevB.85.115130} {\bibfield  {journal} {\bibinfo
  {journal} {Phys. Rev. B}\ }\textbf {\bibinfo {volume} {85}},\ \bibinfo
  {pages} {115130} (\bibinfo {year} {2012})}\BibitemShut {NoStop}%
\bibitem [{\citenamefont {Claeys}\ \emph {et~al.}(2015)\citenamefont {Claeys},
  \citenamefont {De~Baerdemacker}, \citenamefont {Van~Raemdonck},\ and\
  \citenamefont {Van~Neck}}]{claeys15}%
  \BibitemOpen
  \bibfield  {author} {\bibinfo {author} {\bibfnamefont {P.~W.}\ \bibnamefont
  {Claeys}}, \bibinfo {author} {\bibfnamefont {S.}~\bibnamefont
  {De~Baerdemacker}}, \bibinfo {author} {\bibfnamefont {M.}~\bibnamefont
  {Van~Raemdonck}},\ and\ \bibinfo {author} {\bibfnamefont {D.}~\bibnamefont
  {Van~Neck}},\ }\bibfield  {title} {\bibinfo {title} {{Eigenvalue-based method
  and form-factor determinant representations for integrable XXZ
  Richardson-Gaudin models}},\ }\href
  {https://doi.org/10.1103/PhysRevB.91.155102} {\bibfield  {journal} {\bibinfo
  {journal} {Phys. Rev. B}\ }\textbf {\bibinfo {volume} {91}},\ \bibinfo
  {pages} {155102} (\bibinfo {year} {2015})}\BibitemShut {NoStop}%
\bibitem [{\citenamefont {Resare}(2022)}]{resare22}%
  \BibitemOpen
  \bibfield  {author} {\bibinfo {author} {\bibfnamefont {F.}~\bibnamefont
  {Resare}},\ }\emph {\bibinfo {title} {{Richardson models for mesoscopic
  pairing interactions}}},\ \href {https://hdl.handle.net/20.500.12380/305255}
  {Master's thesis},\ \bibinfo  {school} {Chalmers University of Technology}
  (\bibinfo {year} {2022})\BibitemShut {NoStop}%
\bibitem [{\citenamefont {Lee}(2007)}]{lee07}%
  \BibitemOpen
  \bibfield  {author} {\bibinfo {author} {\bibfnamefont {D.}~\bibnamefont
  {Lee}},\ }\bibfield  {title} {\bibinfo {title} {{Spectral Convexity for
  Attractive $\mathrm{SU}(2N)$ Fermions}},\ }\href
  {https://link.aps.org/doi/10.1103/PhysRevLett.98.182501} {\bibfield
  {journal} {\bibinfo  {journal} {Phys. Rev. Lett.}\ }\textbf {\bibinfo
  {volume} {98}},\ \bibinfo {pages} {182501} (\bibinfo {year}
  {2007})}\BibitemShut {NoStop}%
\bibitem [{\citenamefont {Rom\'an}\ \emph {et~al.}(2003)\citenamefont
  {Rom\'an}, \citenamefont {Sierra},\ and\ \citenamefont {Dukelsky}}]{roman03}%
  \BibitemOpen
  \bibfield  {author} {\bibinfo {author} {\bibfnamefont {J.~M.}\ \bibnamefont
  {Rom\'an}}, \bibinfo {author} {\bibfnamefont {G.}~\bibnamefont {Sierra}},\
  and\ \bibinfo {author} {\bibfnamefont {J.}~\bibnamefont {Dukelsky}},\
  }\bibfield  {title} {\bibinfo {title} {{Elementary excitations of the BCS
  model in the canonical ensemble}},\ }\href
  {https://doi.org/10.1103/PhysRevB.67.064510} {\bibfield  {journal} {\bibinfo
  {journal} {Phys. Rev. B}\ }\textbf {\bibinfo {volume} {67}},\ \bibinfo
  {pages} {064510} (\bibinfo {year} {2003})}\BibitemShut {NoStop}%
\bibitem [{\citenamefont {Cheng}\ \emph {et~al.}(2015)\citenamefont {Cheng},
  \citenamefont {Tomczyk}, \citenamefont {Lu}, \citenamefont {Veazey},
  \citenamefont {Huang}, \citenamefont {Irvin}, \citenamefont {Ryu},
  \citenamefont {Lee}, \citenamefont {Eom}, \citenamefont {Hellberg},\ and\
  \citenamefont {Levy}}]{cheng15}%
  \BibitemOpen
  \bibfield  {author} {\bibinfo {author} {\bibfnamefont {G.}~\bibnamefont
  {Cheng}}, \bibinfo {author} {\bibfnamefont {M.}~\bibnamefont {Tomczyk}},
  \bibinfo {author} {\bibfnamefont {S.}~\bibnamefont {Lu}}, \bibinfo {author}
  {\bibfnamefont {J.~P.}\ \bibnamefont {Veazey}}, \bibinfo {author}
  {\bibfnamefont {M.}~\bibnamefont {Huang}}, \bibinfo {author} {\bibfnamefont
  {P.}~\bibnamefont {Irvin}}, \bibinfo {author} {\bibfnamefont
  {S.}~\bibnamefont {Ryu}}, \bibinfo {author} {\bibfnamefont {H.}~\bibnamefont
  {Lee}}, \bibinfo {author} {\bibfnamefont {C.-B.}\ \bibnamefont {Eom}},
  \bibinfo {author} {\bibfnamefont {C.~S.}\ \bibnamefont {Hellberg}},\ and\
  \bibinfo {author} {\bibfnamefont {J.}~\bibnamefont {Levy}},\ }\bibfield
  {title} {\bibinfo {title} {{Electron Pairing Without Superconductivity}},\
  }\href {http://dx.doi.org/10.1038/nature14398} {\bibfield  {journal}
  {\bibinfo  {journal} {Nature (London)}\ }\textbf {\bibinfo {volume} {521}},\
  \bibinfo {pages} {196} (\bibinfo {year} {2015})}\BibitemShut {NoStop}%
\bibitem [{\citenamefont {Bruun}(2002)}]{bruun02}%
  \BibitemOpen
  \bibfield  {author} {\bibinfo {author} {\bibfnamefont {G.~M.}\ \bibnamefont
  {Bruun}},\ }\bibfield  {title} {\bibinfo {title} {{Low-Energy Monopole Modes
  of a Trapped Atomic Fermi Gas}},\ }\href
  {https://doi.org/10.1103/PhysRevLett.89.263002} {\bibfield  {journal}
  {\bibinfo  {journal} {Phys. Rev. Lett.}\ }\textbf {\bibinfo {volume} {89}},\
  \bibinfo {pages} {263002} (\bibinfo {year} {2002})}\BibitemShut {NoStop}%
\bibitem [{\citenamefont {Larkin}\ and\ \citenamefont
  {Varlamov}(2005)}]{larkin05}%
  \BibitemOpen
  \bibfield  {author} {\bibinfo {author} {\bibfnamefont {A.}~\bibnamefont
  {Larkin}}\ and\ \bibinfo {author} {\bibfnamefont {A.}~\bibnamefont
  {Varlamov}},\ }\href@noop {} {\emph {\bibinfo {title} {{Theory of
  Fluctuations in Superconductors}}}}\ (\bibinfo  {publisher} {Oxford
  University Press (Oxford)},\ \bibinfo {year} {2005})\BibitemShut {NoStop}%
\bibitem [{\citenamefont {Cooper}(1956)}]{cooper56}%
  \BibitemOpen
  \bibfield  {author} {\bibinfo {author} {\bibfnamefont {L.~N.}\ \bibnamefont
  {Cooper}},\ }\bibfield  {title} {\bibinfo {title} {{Bound Electron Pairs in a
  Degenerate Fermi Gas}},\ }\href {https://doi.org/10.1103/PhysRev.104.1189}
  {\bibfield  {journal} {\bibinfo  {journal} {Phys. Rev.}\ }\textbf {\bibinfo
  {volume} {104}},\ \bibinfo {pages} {1189} (\bibinfo {year}
  {1956})}\BibitemShut {NoStop}%
\bibitem [{\citenamefont {Zhou}\ \emph {et~al.}(2002)\citenamefont {Zhou},
  \citenamefont {Links}, \citenamefont {McKenzie},\ and\ \citenamefont
  {Gould}}]{zhou02}%
  \BibitemOpen
  \bibfield  {author} {\bibinfo {author} {\bibfnamefont {H.-Q.}\ \bibnamefont
  {Zhou}}, \bibinfo {author} {\bibfnamefont {J.}~\bibnamefont {Links}},
  \bibinfo {author} {\bibfnamefont {R.~H.}\ \bibnamefont {McKenzie}},\ and\
  \bibinfo {author} {\bibfnamefont {M.~D.}\ \bibnamefont {Gould}},\ }\bibfield
  {title} {\bibinfo {title} {{Superconducting correlations in metallic
  nanoparticles: Exact solution of the BCS model by the algebraic Bethe
  ansatz}},\ }\href {https://doi.org/10.1103/PhysRevB.65.060502} {\bibfield
  {journal} {\bibinfo  {journal} {Phys. Rev. B}\ }\textbf {\bibinfo {volume}
  {65}},\ \bibinfo {pages} {060502} (\bibinfo {year} {2002})}\BibitemShut
  {NoStop}%
\bibitem [{\citenamefont {Faribault}\ \emph {et~al.}(2008)\citenamefont
  {Faribault}, \citenamefont {Calabrese},\ and\ \citenamefont
  {Caux}}]{faribault08}%
  \BibitemOpen
  \bibfield  {author} {\bibinfo {author} {\bibfnamefont {A.}~\bibnamefont
  {Faribault}}, \bibinfo {author} {\bibfnamefont {P.}~\bibnamefont
  {Calabrese}},\ and\ \bibinfo {author} {\bibfnamefont {J.-S.}\ \bibnamefont
  {Caux}},\ }\bibfield  {title} {\bibinfo {title} {{Exact mesoscopic
  correlation functions of the Richardson pairing model}},\ }\href
  {https://doi.org/10.1103/PhysRevB.77.064503} {\bibfield  {journal} {\bibinfo
  {journal} {Phys. Rev. B}\ }\textbf {\bibinfo {volume} {77}},\ \bibinfo
  {pages} {064503} (\bibinfo {year} {2008})}\BibitemShut {NoStop}%
\end{thebibliography}%

\end{document}